# Frequency-temperature relations of novel cuts of quartz crystals for thickness-shear resonators

LM Zhang, SY Wang, LT Xie, TF Ma, JK Du, J Wang
Piezoelectric Device Laboratory, Ningbo University,
818 Fenghua Road, Ningbo, Zhejiang 315211, CHINA
Email: wangji@nbu.edu.cn

Y-K Yong
Department of Civil and Environmental Engineering,
Rutgers University, 623 Bowser Road,
NJ 08854, USA

*Abstract*—In a recent study, we have reported that there are many novel cuts of quartz crystal exhibiting the highly treasured cubic frequency-temperature relations which are currently shown only with the AT- and SC-cut. Through setting the first- and second-order derivatives of the frequency respect to temperature to zeroes, a family of quartz crystal cuts with different temperatures of zero frequency (turnover temperatures) has been found and examined. It is now possible to fabricate quartz crystal resonators with turnover temperature near its operating temperature to keep the resonator functioning in a lean and more natural state. By selecting a few cuts based on orientations from our study, we analyzed the thickness-shear vibrations of quartz crystal plates to confirm the superior frequency-temperature relations with the theory of incremental thermal field and Mindlin plate equations and presenting comparisons with known AT- and SC-cut to demonstrate that resonators with newly found cuts can also achieve exceptional frequency stability as demanded.

## I. Introduction

Quartz crystal resonators have frequency and temperature characteristics which are strongly dependent on crystal orientations. In a previous work [1], we obtained a series of quartz crystal cuts with good frequency-temperature relationship through establishing the equations of vibration of plates in the thermal field and setting their first- and second-order derivatives of frequency to vanish. Related to the thermal effects of quartz crystal resonators, Wang et al. studied the modification of Mindlin theory of high-order plate equations [2-3]. For the frequency-temperature characteristics of quartz crystal resonators, Lee and Yong proposed the incremental thermal field theory [4]. Lee and Wang studied the frequency-temperature relations of thickness-shear and flexural vibrations of contoured quartz resonators [5].

After finding the cuts with cubic frequency temperature relation, we consider the Bechmann's approach [6], and adopt the Mindlin plate theory in incremental thermal field [4] to analyze some of them for the vibration frequency and mode shapes with thermal consideration.

## II. Vibrations of Infinite Plate in a Thermal Field

For an infinite quartz crystal plate, by solving the eigenvalue equation, we can obtain three roots ($c_i, i = 1,2,3$). Then substituting them into traction-free boundary condition, we find the plate vibration frequencies

$$f_i = \frac{n}{2b}\sqrt{\frac{c_i}{\rho}}, i = 1,2,3, n = 1,3,5 \cdots. \quad (1)$$

where $f_i$ is the frequency, $b$ is the half thickness and $\rho$ is the density.

From above we get $f_1 > f_2 > f_3$, and their vibration modes are designated as A, B and C. Setting the first- and second-order derivatives of the temperature-frequency equations to zeroes, we obtained the curves of the second-order zero temperature coefficient point with orientations at special angles of doubly-rotated quartz crystals as show in Fig. 1. There are four curves for C mode with cubic temperature-frequency curves named as CCC1-4 and one curve for B mode named CCB1. On these curves, resonators can have the same excellent temperature-frequency characteristics as the most commonly used AT- and SC-cut types.

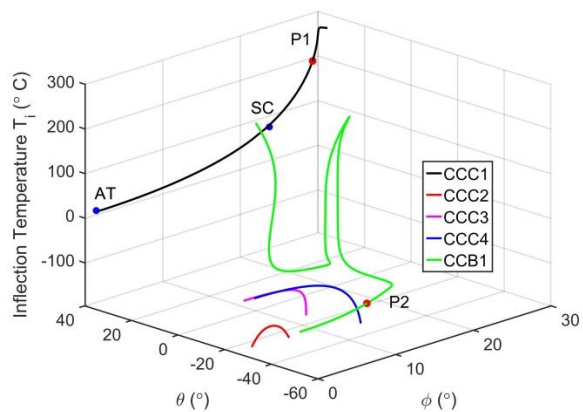

Fig. 1. Three-dimensional curves of $T_f^{(1)} = 0$ and $T_f^{(2)} = 0$ for the thickness mode B and C of a quartz plate as function of the angle $\theta$ and $\varphi$ in rectangular coordinate system.

Since the AT- and SC-cut have been studied extensively for the continuous improvement of resonator products with broad applications, it is just intriguing to us that if all other points on the curves, particularly those in the lower three branches, can also be used to develop resonators with similar or even superior performances. To answer such a question, we have to use the Mindlin plate theory of quartz crystal plates of those orientations to find out the couplings, frequency, and more importantly, the frequency-temperature relations as the preliminary evaluation of possible properties of resonators.

We selected the orientations on the curve with the same $\phi$ angle as the AT- and SC-cut, and the frequency-temperature curve of the corresponding cut shape are plotted and shown in Fig. 2.

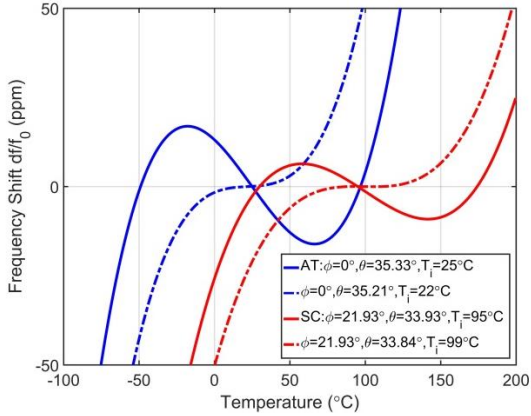

Fig. 2. The frequency-temperature curves of quartz crystal plates of AT- and SC-cut and more orientations with the same $\phi$ angle.

### III. VIBRATIONS OF DOUBLY ROTATED QUARTZ PLATES WITH MINDLIN PLATE EQUATIONS IN AN THERMAL FIELD

The Mindlin plate theory is widely used for thickness-shear vibration analysis of piezoelectric plates. In the Mindlin plate theory, all mechanical displacements and potentials are extended to the power series of plate thickness coordinates [4]

$$u_i(x_1, x_2, x_3, t) = \sum_{n=0}^{\infty} x_2^n u_i^{(n)}(x_1, x_3, t), i = 1,2,3, \quad (2)$$

where $u_i, u_i^{(n)}, x_1, x_2, x_3$ and $t$ are displacements, $n$th-order displacements, coordinates of length, thickness, width and time, respectively.

With the incremental thermal field theory, the first-order equations of motion of the Mindlin plate theory are [4]

$$\beta_{ik} t_{kj,j}^{(0)} = 2b\rho \ddot{u}_i^{(0)}, \ \beta_{ik} t_{kj,j}^{(1)} - \beta_{ik} t_{k2}^{(0)} = \frac{2}{3} b^3 \rho \ddot{u}_i^{(1)}, \quad (3)$$

where $t_{kj}^{(n)}$ are the $n$th-order stress and $\beta_{ik}$ are

$$\beta_{ik} = \delta_{ik} + \alpha_{ik}^{\Theta}, \alpha_{ik}^{\Theta} = \alpha_{ik}^{(1)} \Theta + \alpha_{ik}^{(2)} \Theta^2 + \alpha_{ik}^{(3)} \Theta^3, \quad (4)$$

$$\Theta = T - T_0, i, k = 1,2,3,$$

where $\delta_{ik}, \alpha_{ik}^{(n)}, \Theta$ and $T_0$ are the Kronecker delta, the $n$th-order expansion coefficients [4, 6], and the increment of temperature from the reference temperature $T_0$ ($T_0$=25°C).

Strain-displacement relations are

$$e_{ij}^{(0)} = \frac{1}{2} \left( \beta_{kj} u_{k,i}^{(0)} + \beta_{ki} u_{k,j}^{(0)} + \delta_{i2} \beta_{kj} u_k^{(1)} + \delta_{2j} \beta_{ki} u_k^{(1)} \right),$$

(5)

$$e_{ij}^{(1)} = \frac{1}{2} \left( \beta_{kj} u_{k,i}^{(1)} + \beta_{ki} u_{k,j}^{(1)} \right), i, j, k, l = 1,2,3.$$

Stress-strain relations are

$$t_{ij}^{(0)} = 2b k_{ij}^{(0)} k_{kl}^{(0)} D_{ijkl} e_{kl}^{(0)},$$

(6)

$$t_{ij}^{(1)} = \frac{2}{3} b^3 D_{ijkl}^{(1)} e_{kl}^{(1)}, i, j, k, l = 1,2,3,$$

where $D_{ijkl}$ and $D_{ijkl}^{(1)}$ are the components of thermal elastic constants and modified thermal elastic constants

$$D_{ijkl} = C_{ijkl} + D_{ijkl}^{(1)} \Theta + D_{ijkl}^{(2)} \Theta^2 + D_{ijkl}^{(3)} \Theta^3,$$

(7)

$$D_{ijkl}^{(1)} = D_{ijkl} - \frac{D_{ij22} D_{22kl}}{D_{2222}},$$

where $C_{ijkl}$ are elastic constants and $k_{ij}^{(0)} (i, j = 1,2,3)$ are correction factors [2-3]

$$k_{11}^{(0)} = k_{33}^{(0)} = k_{13}^{(0)} = 1, k_{22}^{(0)} = k_{23}^{(0)} = k_{12}^{(0)} = \sqrt{\frac{\pi^2}{12}}. \quad (8)$$

Substituting (5) and (6) into the equation of motion (3) will give the equation of motion represented by displacements

When using the two-dimensional Mindlin plate theory to analyze high-frequency vibrations of a quartz crystal plate, in order to obtain a solution of closed form, we consider that the acoustic wave propagates only along the length or width of the quartz crystal plate. In order to calculate the dispersion relations of the high-frequency vibrations when waves in quartz crystal plate propagates along the length direction of the plate, we assume that the high-order displacement solutions as

$$u_j^{(0)} = A_j \sin(\eta x_1) e^{i\omega t},$$

(9)

$$u_j^{(1)} = \frac{A_{(j+3)}}{b} \cos(\eta x_1) e^{i\omega t}, j = 1,2,3,$$

where $A_i (i = 1,2,3,4,5,6), \eta, \omega$ and $t$ are the components of the amplitude, wavenumber, coordinates of the plate length or width, angular frequency, and time, respectively.

The normalized parameters are

$$\Omega = \frac{\omega}{\omega_0}, \omega_0 = \frac{\pi}{2b} \sqrt{\frac{c_{66}}{\rho}}, Z = \frac{\eta}{\frac{\pi}{2b}}, D_{pq} = \frac{D_{pq}}{D_{66}}. \quad (10)$$

Substituting equations (9) and (10) into the equations of motion in displacements, the homogeneous equations about the amplitude are obtained as

$$\boldsymbol{MA} = \boldsymbol{0}, \tag{11}$$

where $\boldsymbol{M}$ and $\boldsymbol{A}$ are coefficient and amplitude matrices.

By making the determinant of coefficients of the homogeneous equation group equal to zero, the dispersion equation of the high-frequency vibration of the double-rotated quartz crystal plate in the incremental thermal field can be obtained as [4, 5]

$$F_1[\Omega, Z, D_{pq}(T,\phi,\theta), \beta_{ij}(T,\phi,\theta)] = |\boldsymbol{M}| = 0. \tag{12}$$

The traction-free boundary conditions at $x_1 = \pm a$ are

$$P_1^{(n)} = \beta_{11} t_{11}^{(n)} = 0,$$

$$P_2^{(n)} = \beta_{22} t_{12}^{(n)} + \beta_{23} t_{23}^{(n)} = 0, \tag{13}$$

$$P_3^{(n)} = \beta_{23} t_{12}^{(n)} + \beta_{33} t_{23}^{(n)} = 0, n = 0,1.$$

To satisfy the boundary conditions, from (9) we let

$$u_j^{(0)} = \sum_{r=1}^{6}\left[\alpha_{jr} A_{6r} \sin\left(\frac{Z_r \pi}{2}\frac{x_1}{b}\right)\right]e^{i\omega t},$$

$$u_j^{(1)} = \sum_{r=1}^{6}\left[\alpha_{(j+3)r}\frac{A_{6r}}{b}\cos\left(\frac{Z_r \pi}{2}\frac{x_1}{b}\right)\right]e^{i\omega t}, \tag{14}$$

$$\alpha_{sr} = \frac{A_{sr}}{A_{6r}}, s,r = 1,2,3,4,5,6, j = 1,2,3.$$

Substituting (16) into (15), we can obtain [4, 5]

$$\boldsymbol{NA} = \boldsymbol{0}, \tag{15}$$

where $\boldsymbol{N}$ and $\boldsymbol{A}$ are coefficient and amplitude matrices.

The frequency spectrum equation or the frequency-temperature equation of high frequency vibrations of the quartz crystal plate can be obtained as [4, 5]

$$F_2\left\{Z_r[\Omega, D_{pq}(T,\phi,\theta), \beta_{ij}(T,\phi,\theta)], \alpha_{sr}(\Omega, Z_r), \frac{a}{b}\right\} = |\boldsymbol{N}| = 0. \tag{16}$$

Now we have complete formulation of high frequency vibrations of quartz crystal plates of novel orientations in a thermal field with the Mindlin plate equations. For straight-crested waves in the length direction, we can obtain the frequency spectra to evaluate the frequency-temperature properties for resonators with optimal performance. We have more orientations for quartz crystal resonators in addition to known AT- and SC-cut types.

## IV. NUMERICAL EXPLAMES

We choose two cases from Fig. 1 such as $P_1$ and $P_2$ to analyze the frequency spectrum, mode shapes, and frequency-temperature relations by using the first-order Mindlin plate equations in a thermal field.

The first point $P_1$ on the line CCC1 belongs to mode C with $\phi = 27.1°, \theta = 32.6°$ and $T_0 = 216.84\,°C$ which has a higher inflection point temperature. Frequency spectra of plates with this orientation for $x_1$ direction propagation waves in an quartz plate are calculated from (16) and shown in Fig. 3. The normalized displacement mode shapes at resonances of predominantly fundamental thickness-shear vibrations are plotted as shown in Fig. 4.

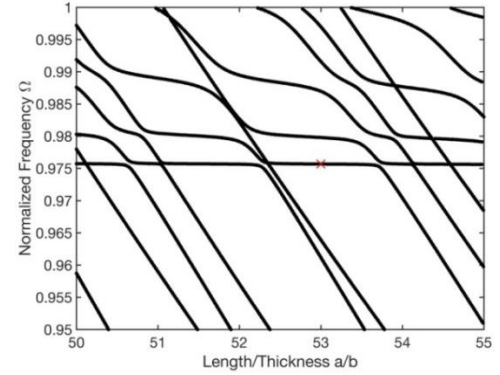

Fig. 3. Frequency spectra of a doubly-rotated cut of quartz crysta plate at $P_1$ ($\phi = 27.1°, \theta = 32.6°, T_0 = 216.84°C$)

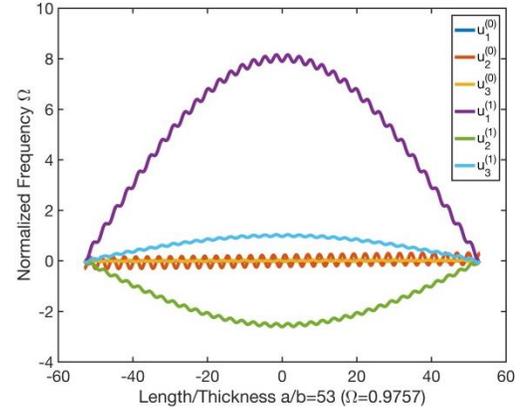

Fig. 4. Normalized displacements of a doubly-rotated cut of quartz crystal plate at $P_1$ ($\phi = 27.1°, \theta = 32.6°, T_0 = 216.84°C$).

To calculate the frequency-temperature behavior, with the length-thickness ratio $a/b =53$, from (16) we can obtain the change in resonance frequencies as a function of temperature for infinite and finite quartz plates and results are shown in Fig. 5.

Select the second point $P_2$ on the line CCB1 belongs to mode B with $\phi =10.2°, \theta =$ -47.0153° and $T_0 =$ -104°C, and the inflection point temperature on the branch is below zero. Frequency spectra curves of plates with this orientation are shown in Fig. 6. At the low temperature point, we found that the normalized displacement mode shapes at resonances of predominantly fundamental thickness-shear vibrations are all shifted to the B mode as show in Fig. 7.

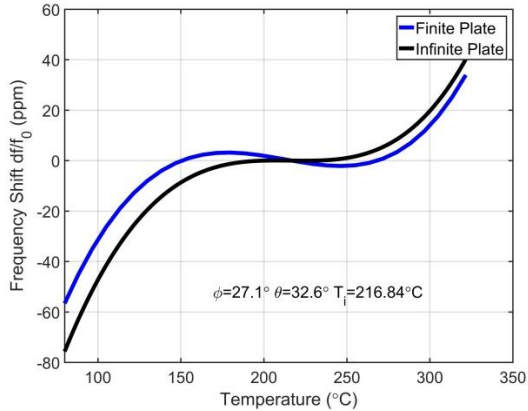

Fig. 5. Frequency-temperature curves of thickness-shear vibrations at $P_1$

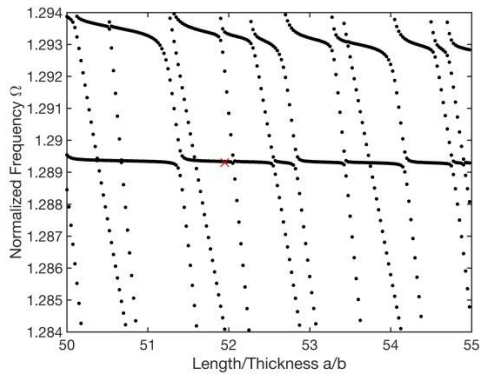

Fig. 6. Frequency spectra of a doubly-rotated quartz crystal plate at $P_2$ ($\phi = 10.2°, \theta = -47.0153°, T_0 = -104°C$)

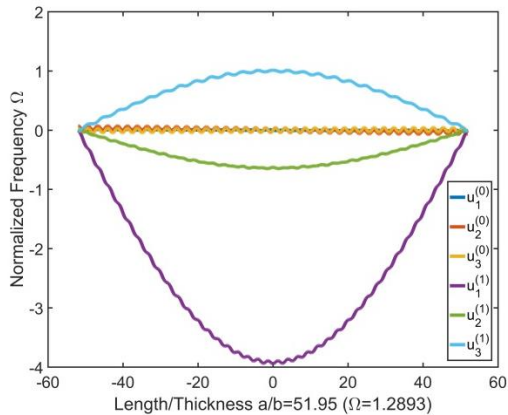

Fig.7. Normalized displacements of a doubly-rotated quartz crystal plate at $P_2$ ($\phi = 10.2°, \theta = -47.0153°, T_0 = -104°C$)

The frequency-temperature relations for infinite and finite quartz plates are shown in Fig. 8. It can be seen that the frequency-temperature curve of the negative branch has a completely different direction from the positive frequency-temperature curve as shown in Fig. 2.

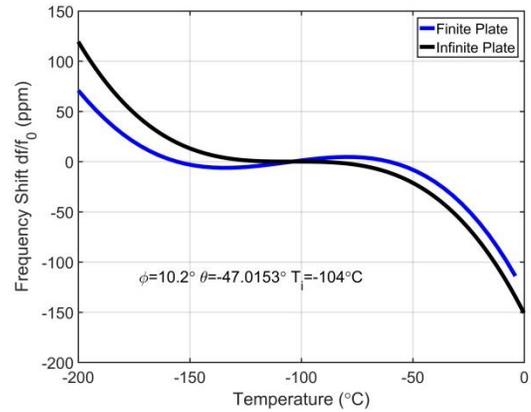

Fig. 8. Frequency-temperature curves of thickness-shear vibrations at $P_2$

## V. CONCLUSIONS

The Mindlin plate theory in an incremental thermal field has been successfully adopted for the analysis of high frequency vibrations of quartz crystal plates of special doubly-rotated cuts, and the frequency-temperature characteristics are excellent. We have also found that the negative branch of quartz crystal cuts may have completely different vibration modes in comparison with the positive branch, and the negative branch of quartz crystal cuts have elevated thickness-shear frequency. The analysis of different cuts can help us finding quartz crystal resonators suitable for extreme operating environments. The procedure presented here is adequate for the evaluation of the novel cuts and optimal design of new types of quartz crystal resonators.

ACKNOWLEDGMENT

This research is supported by the National Natural Science Foundation of China (Grant Nos. 11372145, 11672142, & 11772163).